\documentclass[twocolumn,superscriptaddress,floatfix,preprintnumbers,prl ,nofootinbib,hyperref,longbibliography]{revtex4-1}
\usepackage[utf8]{inputenc}
\usepackage{mathrsfs}
\usepackage{graphicx}
\usepackage{amssymb}
\usepackage{textcomp}
\usepackage{tabularx}
\usepackage{dcolumn}
\usepackage{bm}
\usepackage{times}
\usepackage{color}
\usepackage{slashed}
\usepackage{multirow}
\usepackage{verbatim}
\usepackage{cancel}
\usepackage{subfigure}
\usepackage{ulem}

\usepackage{stackrel}
\usepackage{pgfplots}
\usetikzlibrary{positioning}
\usetikzlibrary{snakes}

\usepackage[colorlinks=true, pdfstartview=FitV, linkcolor=blue, citecolor=blue, urlcolor=blue]{hyperref}
\def\lsim{\mathrel{\raise.3ex\hbox{$<$\kern-.75em\lower1ex\hbox{$\sim$}}}}
\def\gsim{\mathrel{\raise.3ex\hbox{$>$\kern-.75em\lower1ex\hbox{$\sim$}}}}

\def\beq{\begin{equation}}
\def\eeq{\end{equation}}
\def\bea{\begin{eqnarray}}
\def\eea{\end{eqnarray}}

\def \({\left(}
\def \){\right)}
\def \[{\left[}
\def \]{\right]}
\def \l|{\left|}
\def \r|{\right|}

\def \bea {\begin{eqnarray}}
\def \eea {\end{eqnarray}}

\definecolor{orange}{rgb}{1,0.5,0}

\begin{document}
\bibliographystyle{apsrev4-1}

\title{Oscillations of Ultralight Dark Photon into Gravitational Waves}

\author{Wei Chao}
\email{chaowei@bnu.edu.cn}
\author{Jing-jing Feng}
\email{fengjj@mail.bnu.edu.cn}
\affiliation{
Center for Advanced Quantum Studies, Department of Physics, Beijing Normal University, Beijing, 100875, China
}
\author{Huai-ke Guo}
\email{guohuaike@ucas.ac.cn}
\affiliation{
International Centre for Theoretical Physics Asia-Pacific,
University of Chinese Academy of Sciences, 100190 Beijing, China
}
\author{Tong Li}
\email{litong@nankai.edu.cn}
\affiliation{
School of Physics, Nankai University, Tianjin 300071, China 
}

\vspace{3cm}

\begin{abstract}
The discovery of gravitational waves~(GWs) opens a new window for exploring the physics of the early universe. Identifying the source of GWs and their spectra today turns out to be the important tasks so as to assist the experimental detection of stochastic GWs. 
In this paper, we investigate the oscillations of the ultralight dark photon (ULDP) into GWs in the dark halo. Assuming dark matter is composed of the ULDP and there are primordial dark magnetic fields (PDMFs) arising from the axion inflation and/or the dark phase transition, then the ULDP can oscillate into the GW when it passes through an environment of PDMFs.   
We derive the local energy density of GWs in the galaxy cluster induced by the instaneous oscillation of ULDP in the PDMFs. These stochastic local GWs exhibit a pulse-like spectrum, with frequency depending on the mass of the ULDP, and can be detected in Pulsar Timing Arrays (PTAs) or future space-based interferometers.
We also find that the low-frequency GW signal observed by the NANOGrav collaboration and other PTA experiments can be explained by the oscillation of the ULDP in the PDMFs in the early universe.
\end{abstract}

\maketitle
%%%%%%%%%%%%%%%%%%%%%%%%%%%%%%%%%%%
\section{Introduction}
\label{sec:Intro}
%%%%%%%%%%%%%%%%%%%%%%%%%%%%%%%%%%%

The discovery of gravitational waves (GWs) from the merger of two astrophysical objects by the LIGO and VIRGO collaborations~\cite{LIGOScientific:2016aoc,LIGOScientific:2016sjg,LIGOScientific:2017bnn,LIGOScientific:2017ycc,LIGOScientific:2017vox,LIGOScientific:2020zkf,LIGOScientific:2020iuh} has led to a new era for cosmology, astrophysics and high energy physics. 
%The GWs of black holes and neutron stars are at frequencies $f\gtrsim 10$ Hz. 
Analogous to electromagnetic radiation, there should be GWs in various frequency bands in the universe. They carry the information of the universe in various different eras and may revolutionize our understanding of the universe. 
%In the past several decades, 
%many devices were designed for detecting the GWs.  A traditional device is the interferometer. The central focus of 
In the past several decades, the ground-based interferometers have been designed to detect the GWs with hertz to kilohertz frequency, which may detect the GWs of black holes and neutron stars. The space-based interferometers operate to search for GWs in the frequency range of $0.01~{\rm Hz} < f < 1~{\rm Hz}$~\cite{Caprini:2019egz,Ruan:2018tsw}, which may detect stochastic GWs from phase transitions and other exotic processes. Alternatively, the pulsar time arrays (PTAs) are proposed to detect nanohertz GWs inspired by the orbital decay of the binary pulsar in 1970s~\cite{Hulse:1974eb}. They look for the GW effects of incoming electromagnetic signals from an array of 20–50 well-known millisecond pulsars.

More recently, several PTAs including NANOGrav~\cite{NANOGrav:2023gor,NANOGrav:2023hvm,NANOGrav:2023hfp,NANOGrav:2023icp,NANOGrav:2023ctt,NANOGrav:2023hde,NANOGrav:2023tcn}, European PTA~\cite{EPTA:2023fyk}, Parkes PTA~\cite{Zic:2023gta} and China PTA~\cite{Xu:2023wog} show the first evidence of the nHz stochastic gravitational wave backgrounds (SGWBs) of our universe. 
%Understanding the origin of these SGWs is an important issue as they may origin from very different fundamental physics. 
%These SGWBs may origin from very different fundamental physics.
It has been shown that these nHZ SGWs may come from either supermassive black hole binaries~\cite{NANOGrav:2023hfp} or exotic new physics~\cite{NANOGrav:2023hvm}, such as first order phase transitions~\cite{Fujikura:2023lkn,Athron:2023xlk,Addazi:2023jvg,Jiang:2023qbm,Xiao:2023dbb,Abe:2023yrw,Gouttenoire:2023bqy}, topological defects like cosmic string~\cite{Jiao:2023wcn,Kitajima:2023vre,Wang:2023len,Eichhorn:2023gat,Yamada:2023thl,Ahmed:2023pjl} or domain walls~\cite{Lu:2023mcz,Bai:2023cqj,Gouttenoire:2023ftk,Kitajima:2023cek}, scalar-induced gravitational waves~\cite{Cai:2023dls,Wang:2023sij} and etc. For more exotic model buildings, we refer the readers to Refs.~\cite{NANOGrav:2023gor,NANOGrav:2023hvm,NANOGrav:2023hfp,NANOGrav:2023icp,NANOGrav:2023ctt,NANOGrav:2023hde,NANOGrav:2023tcn,EPTA:2023fyk,Zic:2023gta,NANOGrav:2023hfp,NANOGrav:2023hvm,Fujikura:2023lkn,Athron:2023xlk,Addazi:2023jvg,Jiang:2023qbm,Xiao:2023dbb,Abe:2023yrw,Gouttenoire:2023bqy,Jiao:2023wcn,Kitajima:2023vre,Wang:2023len,Eichhorn:2023gat,Yamada:2023thl,Ahmed:2023pjl,Lu:2023mcz,Bai:2023cqj,Gouttenoire:2023ftk,Kitajima:2023cek,Cai:2023dls,Wang:2023sij,Xu:2023wog} and the references cited therein for details.

In this paper, we propose a new mechanism of generating the SGWB via the oscillation of ultralight dark matter (ULDM) in the primordial magnetic field (PMF) or in the primordial dark magnetic field (PDMF). Similar to PMFs, the PDMFs may be generated in the early universe during inflation~\cite{Nakai:2020cfw}, dark sector phase transitions~\cite{Nakai:2020oit} and(or) recombination~\cite{Sethi:2004pe}. ULDMs are appealing dark matter (DM) candidates and may solve astrophysical problems faced by the cold DM. The problems include but not limited to the disagreement between cold DM predictions and the observed small-scale structures~\cite{Hu:2000ke,DelPopolo:2016emo}, the missing satellite problem~\cite{Kim:2017iwr} and the core–cusp problem~\cite{deBlok:2009sp} in dwarf galaxies. As an illustration of the mechanism, we take the ULDM as the ultralight dark photon (ULDP)~\cite{Fabbrichesi:2020wbt} that mixes with photon via the kinetic term. Given the conversion rate of ULDP into GW per second in the cluster, we can estimate the local energy density of stochastic GWs produced from the oscillations of ULDP in PDMFs. It shows a stochastic signal with fixed frequency that depends on the mass of the ULDP. We show that this signal is detectable in PTAs or future space-based interferometers. We further show that the oscillation of ULDP in the PDMF in the early universe can also produce a band signal of the stochastic GWs caused by the red-shift. This scenario can address the observed signal by PTAs. Alternatively, the observed result of the PTAs as well as the effective number of neutrinos may put constraint on the strength of the PDMF as well as the dark photon mass.

The rest of this paper is organized as follows. In Sec.~\ref{sec:ULDP} we give the analytical calculation of ULDP oscillations in PDMF. The local energy density of stochastic GWs as well as the GW flux are derived. Sec.~\ref{sec:earlyU} is devoted to address the NANOGrav results via ULDP oscillations in  PDMF in the early universe. Finally, in Sec.~\ref{sec:Con} we draw our main conclusions.

%%%%%%%%%%%%%%%%%%%%%%%%%%%%%%%%%%%%%%%%%%%%%%
\section{The ULDP oscillations and the local energy density of stochastic GWs}
\label{sec:ULDP}
%%%%%%%%%%%%%%%%%%%%%%%%%%%%%%%%%%%%%%%%%%%%%%

To rationalize the  oscillation of ULDP, it is necessary to start with the total action for the ULDP. In general, the action for a Lagrangian density $\cal L $ that minimally couples to gravity is $S=\int d^4 x \sqrt{-g}{\cal L}$, where $ g $ is the determinant of the background metric $g_{\mu\nu}^{}$. The action of ULDP quantum electrodynamics (QED) can be written as
\begin{small}
\bea
S= \int d^4 x \sqrt{-g } \left({1\over 2} M_{pl}^2 R -{1\over 4}\hat  F_{\mu \nu } g^{\alpha \mu }_{} \hat F_{\alpha \beta}^{} g^{\beta \nu}_{} \right.\nonumber \\ \left.- {\varepsilon\over 2} F_{\mu \nu } g^{\alpha \mu }_{} \hat F_{\alpha \beta}^{} g^{\beta \nu}_{} \right)  \;, 
\label{eq:action}
\eea
\end{small}where $R$ is the Recci scalar, $M_{pl}=2.4\times 10^{18}~{\rm GeV}$ is the reduced Planck mass, the field strength tensors are $ F_{\mu\nu}^{} \equiv \partial_\mu^{} A_\nu^{}-\partial_\nu^{} A_\mu^{}$  and $ \hat F_{\mu\nu}^{} \equiv \partial_\mu^{} A_\nu^{\prime}-\partial_\nu^{} A_\mu^{\prime}$ with $A^{}_\mu$ the electromagnetic field and $A^{\prime}_\mu$ the dark photon field, $g_{\mu\nu}^{} =\eta_{\mu\nu}^{} + h_{\mu\nu}^{}$ with $\eta_{\mu\nu}$ being the metric in flat spacetime  
%\TL{(1,-1,-1,-1) or (-1,1,1,1)?} 
and $|h_{\mu\nu}| \ll 1$, and $\varepsilon$ is a small parameter describing the kinetic mixing between the photon and the dark photon. We have neglected the Euler-Heisenberg term from the vacuum polarization~\cite{Heisenberg:1936nmg} for simplicity. The equation of motion (EOM) for the ULDP is
\begin{small}
\bea
\partial_\nu \hat F^{\mu\nu} = \hat j^\mu + \partial_\nu^{} \left( -{1\over 2} h \hat F^{\mu\nu } + \hat F^{\mu\alpha} h^\nu_\alpha -\hat F^{\nu \alpha } h^{\mu }_\alpha \right)&&\nonumber \\
+\varepsilon \partial_\nu^{} \left( -{1\over 2} h F^{\mu\nu } + F^{\mu\alpha} h^\nu_\alpha -F^{\nu \alpha } h^{\mu }_\alpha \right)&& 
\eea
\end{small}where $\hat j^\mu$ is the dark photon current in the flat space, and the second and third terms are derived from the interactions in the Eq.~(\ref{eq:action}). For the EOM of the electromagnetic field, we refer the readers to Ref.~\cite{Domcke:2023bat} for details. The Schr\"{o}dinger-like equations relevant for the oscillations of ULDM take the following form~\cite{Raffelt:1987im,Ejlli:2018hke,Liu:2023mll}
\begin{small}
\begin{eqnarray}
i{\partial \over \partial z}\left( \matrix{\gamma_{\parallel} \cr \gamma^\prime_{\parallel} \cr G_{\times}} \right) =\left[ \omega + \left(  \matrix{ \Delta_{\gamma \gamma}^{} & \Delta_{\gamma \gamma^\prime}^{} &\Delta_{\gamma h}^{}  \cr \Delta_{\gamma \gamma^\prime}^{} & \Delta_{\gamma^\prime \gamma^\prime}^{} & \Delta_{\gamma^\prime h }^{}  \cr \Delta_{\gamma h}^{} & \Delta_{\gamma^\prime h}^{} & 0 }\right) \right]\left( \matrix{\gamma_{\parallel} \cr \gamma^\prime_{\parallel} \cr G_{\times}} \right) 
\label{eq:osc}
\end{eqnarray}
\end{small}where $\omega $ is the energy of fields, $\gamma_\parallel$ and $\gamma_{\parallel}^\prime$ are the vector fields parallel to the $B_T$ component, which is transverse to the GW propagation direction with $B$ being the external (dark) magnetic field, and $G_{\times }$ is the cross polarization mode of the GW. The elements in the $3\times 3$ propagation matrix are 
$
\Delta_{\gamma^{} \gamma^{} } =\Delta_{\rm pla}^{} + \Delta_{\rm vac}^{} + \Delta_{\rm CM}^{},  ~
\Delta_{\gamma^\prime \gamma^\prime} \approx  {m_{A^\prime}^2 \over 2 \omega },~
\Delta_{\gamma^{} h} = {1\over 2} \kappa \left| B_T^{} \right |,~ 
\Delta_{\gamma^{\prime} h} = {\varepsilon \over 2} \kappa \left| B_T^{} \right | + {1\over 2} \kappa \left|B_T^\prime \right |,
$
where $\Delta_{\rm pla} = -2\pi \alpha n_c /(\omega m_c )$ with $\alpha$, $n_c$ and $m_c$ being the fine structure constant,
the number density and the invariant mass of charged plasma particles and $\Delta_{\rm vac} = 7 \alpha \omega /(90 \pi)(B_T/B_c)^2$ encode the plasma and QED vacuum effects~\cite{Raffelt:1987im}, respectively, $m_{A^\prime}^{}$ is the mass of the dark photon, $B_T$ ($B_T^\prime$) is the magnitude of the PMF (PDMF) transverse to the GW propagation direction, and $\Delta_{\rm CM} \sim B_T^2 $~\cite{Ejlli:2018hke}.  
$\Delta_{\gamma^{(\prime)} h}$ describes the mixing between (dark) photon and the GW with $\kappa = (16\pi G)^{1/2}$. As $\varepsilon\ll 1$~\cite{Fabbrichesi:2020wbt}, we ignore $\varepsilon$ in the following analysis and $\Delta_{\gamma \gamma^\prime}^{}$ can be approximately set to zero for simplicity. Considering a homogeneous and isotropic ULDP flux, we average over the alignment of the DM velocity to the magnetic field giving $\langle B_T^2 \rangle ={1\over 2}B^2 $.

For convenience, we define the matrix in the square brackets in Eq.~(\ref{eq:osc}) as $\cal H$. It can be diagonalized by the $3\times3$ orthogonal transformation: $ {\cal U}^\dagger  {\cal H} {\cal U}^* =\hat {\cal H} = {\rm diag} \{ \rho_{1}^{},~\rho_2^{},~\rho_3^{} \}$, where ${\cal U}$ is a $3\times 3$ unitary matrix and $\rho_i$ is the eigenvalue. Then, the $i\to j$ oscillation probability becomes
\begin{small}
\bea
P(i\to j ) = \left| \sum_\alpha^{} {\cal U}^*_{i \alpha } e^{-i \rho_\alpha^{}  z } {\cal U}_{j \alpha}^{}   \right|^2 \;,~~i,j\in \gamma,\gamma',G_\times \; . 
\label{master0}
\eea
\end{small}This result is analogous to the case of three-flavor neutrino oscillations. It is difficult to derive the analytical results of the above oscillations.
%It should be mentioned that it is a little bit difficult to derive analytical result of oscillations, just-like the case of three flavor neutrino oscillations. 
Actually, considering that the effect of the $\gamma-\gamma^\prime$ mixing is sub-dominate, one can only focus on two components' oscillation to simplify the calculation. Setting $\Delta_{\gamma\gamma'}\approx 0$, the probability $P(\gamma^\prime \to G_\times )$ can be given as
\begin{small}
\bea
P(\gamma^\prime \to G_\times ) ={4 \Delta_{\gamma^\prime h}^2 \over 4 \Delta_{\gamma^\prime h}^2 + \Delta_{\gamma^\prime \gamma^\prime}^2 } \sin^2 \left( {1\over 2 }\sqrt{ 4 \Delta_{\gamma^\prime h}^2 + \Delta_{\gamma^\prime \gamma^\prime}^2 } z \right)\;.
\eea 
\end{small}which is consistent with the results given in Refs.~\cite{Domcke:2020yzq,Irastorza:2018dyq}. One can see that the ULDP-graviton conversion is enhanced in the regime of coherent magnetic fields. This is very similar to the case of the matter effect in neutrino oscillations~\cite{Wolfenstein:1977ue}. The existence  of magnetic fields in the intra-cluster medium has been estimated by many methods, resulting in the typical magnetic field strength of the order ${\cal O}(B)\sim \mu G$~\cite{Govoni:2004as}. The coherent length of these fields are expected to be in the range of $1-10~{\rm kpc}$.  In this study, we assume that the PDMF is produced either from the axion inflation or the dark phase transition, with the field strength similar to that of the PMF in glaxy cluster.

The ULDP traveling in the PDMF and oscillating into the GW has the following conversion rate per ULDP per second 
\begin{small}
\bea
{\cal R}= \left\{
\begin{array}{l r}
{2 \Delta_{\gamma^\prime h}^2 \over 4 \Delta_{\gamma^\prime h}^2 + \Delta_{\gamma^\prime \gamma^\prime}^2 }{1\over\mathbb{T}}\;,  &  L\gg v \mathbb{T}\;,\\
&\\
 P(\gamma^\prime \to G_\times^{} )  { 1 {\rm kpc} \over L} \times 1.02\times 10^{-14}\;,   &  L \ll v \mathbb{T}\;,
 \end{array}
 \right. 
\label{20231204}
\eea
\end{small}where $v\sim 10^{-3} c$ with $c$ being the speed of light is the velocity of the ULDP, $ \mathbb{T}$ is the oscillation period and $L$ is the ULDP traveling length. The  result in the second line of the Eq.~(\ref{20231204}) is consistent with the formula describing axion-photon oscillations in Ref.~\cite{Conlon:2013txa}.  Summing over the PDMF domain, we obtain the luminosity of the oscillations from ULDP to GW 
\bea
{\cal L} (r)= {\rho_{\rm DM}  } (r)  {\cal R}  ~~{\rm GeV}\cdot {\rm cm}^{-3} \cdot {\rm s}^{-1}\;,
\eea
where $\rho_{\rm DM} =n_{\rm DM}^{} m_{\rm DM}^{} $ is the energy density of the DM in the galaxy clusters (GCs) with $m_{\rm DM}$ and $n_{\rm DM}^{}$ being the DM mass and the DM number density, respectively. Clearly, a profile of DM distribution in GC is needed for simulations. There are several models describing DM profile in the universe, such as Navaro-Frenk-White (NFW) profile~\cite{Navarro:1995iw,Navarro:1996gj}, Einasto profile~\cite{Einasto:1965czb} and Burkert profile~\cite{Burkert:1995yz}, etc. In this paper, we take the generalized NFW profile~\cite{Navarro:1996gj} as the input
\bea
\rho_{\rm NFW } (r) = {M_0 \over 4\pi r_s^3} {1 \over (r/r_s)^\beta (1+ r/ r_s)^{3-\beta}} \;,
\eea  
where $M_0$ is the mass normalization, $r_s $ is the scale radius, and $\beta$ is the characteristic power for the inner part of the potential. 

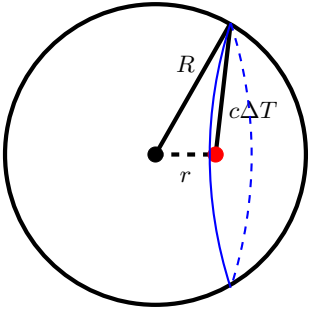
\begin{figure}[t]
\begin{center}
\begin{tikzpicture}
\draw [dashed, ultra thick] (1,1) -- (1.8,1);
\draw [-, ultra thick] (1.8,1) -- (2,2.75);
\draw [-, ultra thick] (1,1) -- (2,2.75);
\draw [fill, red](1.8,1) circle (0.1);
\draw [fill](1,1) circle (0.1);
\draw[style,ultra thick] (1,1) ellipse (2 and 2);
%\draw [-latex,ultra thick] (1,0.5) -- (1.1,0.5);
%\draw [-latex,ultra thick] (1,1.5) -- (1.1,1.5);
\node[black, thick] at (1.4, 0.7) {$r$};
\node[black, thick] at (2.3, 1.6) {$c \Delta T$};
\node[black, thick] at (1.4, 2.2) {$R$};
\draw[blue, thick] (2,2.75) arc(162:198:5.7);
\draw[dashed,blue, thick] (2,2.75) arc(18:-18:5.7);
\end{tikzpicture}
\caption{The illustration for the calculation of the ratio of a graviton which is produced in the GC via oscillations and still stays within the GC after a time interval $\Delta T$. }\label{feynmand}
\end{center}
\end{figure}

Now we can estimate the local energy density of GWs produced from ULDP oscillations in the galaxy cluster. For simplicity, we may assume that a galaxy cluster is a standard spheroid with radius $R \sim 0.63~{\rm Mpc}$~\cite{Conlon:2013txa} and a volume of $V_{\rm GC}\sim 1~{\rm Mpc}^3$. Once the GWs are produced, they fly away the cluster with the speed of light. Thus, we need to estimate the ratio of GWs that still stay in the spheroid of the galaxy cluster in a typical time interval $\Delta T$. 
%which is the time it takes for GW produced at the spherical center fly out the spheroid.  
According to the illustrative plot in Fig.~\ref{feynmand}, the ratio of GWs that are produced at a distance $r$ to the spherical center and still stay within the galaxy cluster in the time interval $\Delta T$ is 
\begin{small}
\bea
f(r, ~\Delta T) =\left \{\begin{array}{crl }
1\;, &  & c \Delta T \leq R- r\;, \\
{1\over 2} \left(  1- { r^2 + c^2 \Delta T^2 - R^2 \over 2 r c \Delta T}\right)\;, &  & c \Delta T > R-r\;,  \\
0\;, & & c \Delta T \geq R+r\;,
\end{array}  \right.
\eea
\end{small}where $R$ is the radius of the galaxy cluster. As a result, the local energy density of GW produced from oscillations at $t_0$ can be written as
\begin{small}
\bea
\rho_{\rm LGW}^{} (t_0)= {1\over V_{\rm GC }} \int_{t_0 - 2T}^{t_0} dt \int_0^R dr  {\cal L}(r) \cdot f(r, t_0- t) 4 \pi r^2\;.
 \label{master1}
\eea
\end{small}The local relic density of GW can be defined as $\Omega_{\rm LGW} = \rho_{\rm LGW}/\rho_C$ with $\rho_C={3\over 8\pi G}H_0^2 =1.05 \times 10^{-5} h^{2} ~{\rm GeV/cm^3}$ being the critical density, where $h=0.67$~\cite{ParticleDataGroup:2022pth}. Note that the real energy density of GW should be slightly larger than that given in Eq.~(\ref{master1}) as the GW produced at the border of the spheroid takes about $2T$ time to fly out of the GC. One may also  estimate the flux of GWs on a specific celestial body using the following formula 
\bea
{\cal D}_{\rm GW} = \int  {d n_{\rm DM} (r) \over d E }{  {\cal R}  d^3 r \over 4 \pi |\vec r -\vec d |^2} \;,
\eea 
where $\vec d$ is the position vector of the  celestial body. Given the NFW profile of the ULDP in the Milky Way, one can estimate the flux of GWs on the earth as ${\cal D}_{\rm GW}^{\rm Earth}=1.72\times 10^{-7} /m_{\rm ULDP}^2 ~{\rm cm}^{-2}\cdot {\rm s}^{-1}\cdot {\rm eV}^{-1}$, which will be useful in detecting GWs on the Earth via GW-photon oscillations.

\begin{figure*}[t]
\centering
\includegraphics[width=0.45\textwidth]{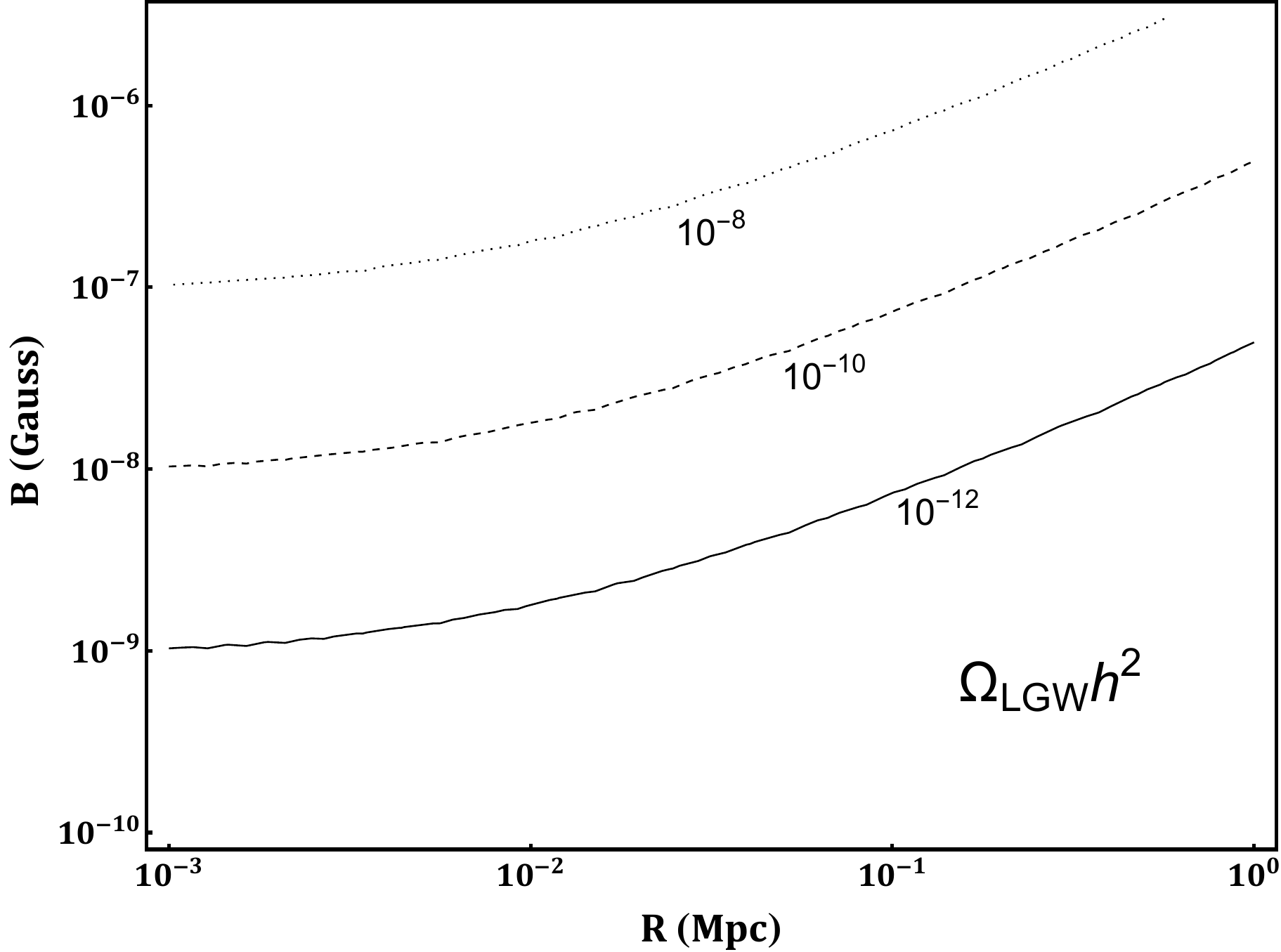}
\quad\quad
\includegraphics[width=0.45\textwidth]{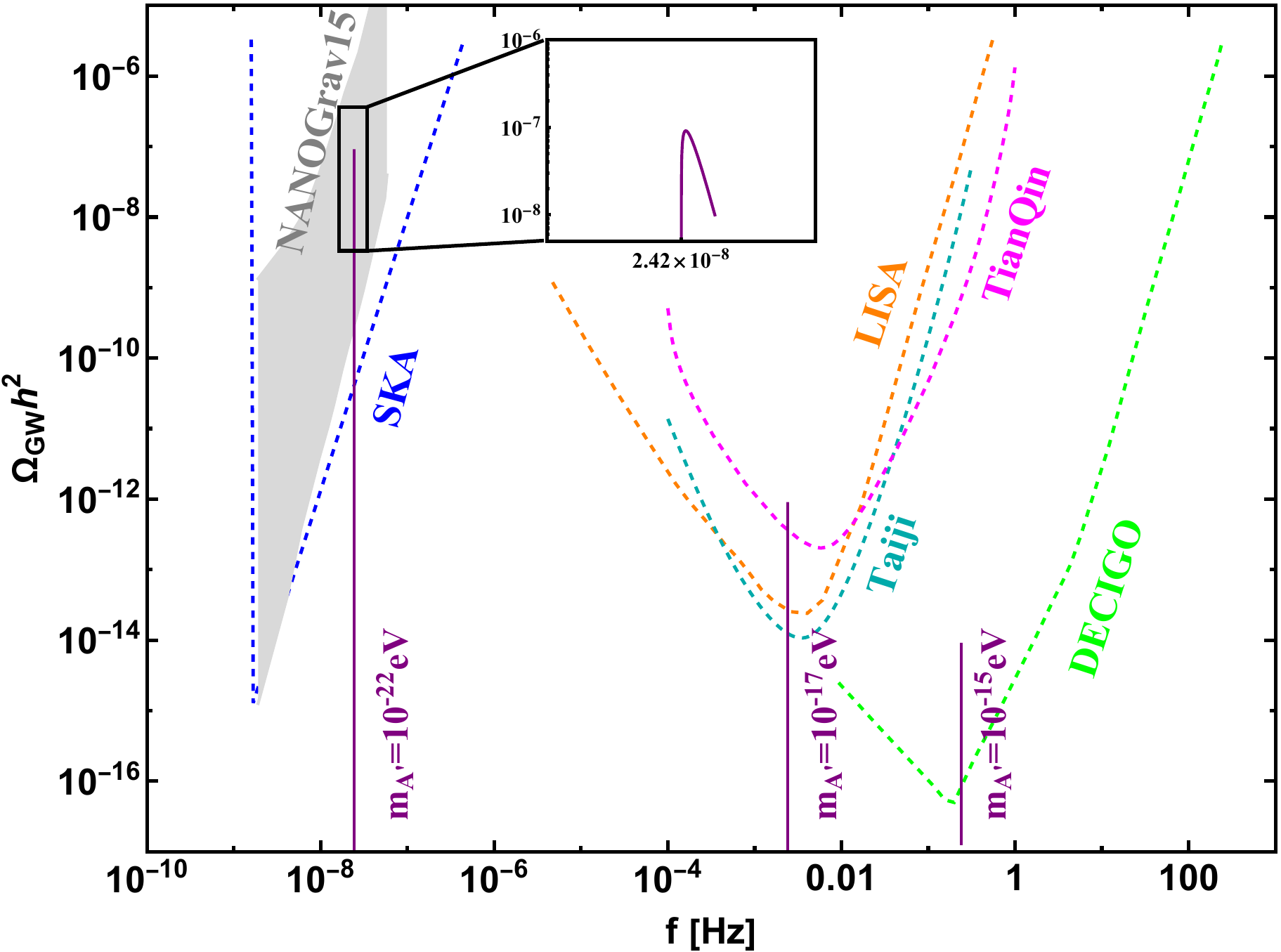}
\caption{Left-panel: Contours of the local relic abundance of the stochastic GWs produced via dark photon oscillations in the dark primordial magnetic field. The curves from top to bottom correspond to $\Omega_{\rm LGW} h^2 = 10^{-8},~10^{-10}$ and $10^{-12}$, respectively. Right-panel: Relic density of local GWs as a function the frequency. The vertical lines from left to right correspond to $m_{A^\prime}=10^{-22}~{\rm eV}$, $10^{-18}~{\rm eV}$ and $10^{-15}~{\rm eV}$, respectively. The gray region represents the signal observed by the NANOGrav collaboration as reference. The blue, orange, green, magenta and dark green dashed curves show the projected sensitivity of the SKA~\cite{Janssen:2014dka}, LISA~\cite{LISA:2017pwj} and DECIGO~\cite{Kawamura:2011zz}, TianQin~\cite{TianQin:2015yph} and Taiji~\cite{Ruan:2018tsw} collaboration, respectively.} \label{fig:2GWRs}
\end{figure*}

Now we consider the spectrum of the local GWs. The frequency of the GW produced from oscillations is $E_{\rm ULDP}/ h$ with $E_{\rm ULDP}$ being the energy of ULDP and $h$ the Planck constant. In the left-panel of Fig.~\ref{fig:2GWRs}, we show the contours of the local abundance of the stochastic GWs produced from the oscillation of ULDP in the presence of PDMF by setting $m_{A^\prime} = 10^{-22} ~{\rm eV}$. The horizontal and vertical axes correspond to the radius of the cluster and the average strength of the PDMF, respectively. The solid curves from top to bottom correspond to $\Omega_{\rm LGW} h^2 = 10^{-8},~10^{-10}$ and $10^{-12}$, respectively. Apparently, these GWs may cause exotic signals in the space-based or ground-based interferometers with the peak frequency depending on the mass of the ULDP. We show the relic density of the local GW as a function of its frequency in the right panel of the Fig.~\ref{fig:2GWRs}. The field strength of the PDMF is set to be $B^\prime \sim 1 ~{\rm \mu G}$, which is in the same order of the strength of the PMF~\cite{Subramanian:2015lua}. The vertical lines from left to right correspond to $m_{A^\prime}=10^{-22}~{\rm eV}$, $10^{-18}~{\rm eV}$ and $10^{-15}~{\rm eV}$, respectively. The typical signature of these local GWs is stochastic but with fixed frequency and can be detected in PTAs or future space-based interferometers with frequency depending on the ULDP mass.

%%%%%%%%%%%%%%%%%%%%%%%%%%%%%%%%%%%%%%%%%%
\section{ULDP Oscillations in the early universe}
\label{sec:earlyU}
%%%%%%%%%%%%%%%%%%%%%%%%%%%%%%%%%%%%%%%%%%

Similar to the generation mechanism of the primordial magnetic fields~\cite{Subramanian:2015lua}, PDMFs could be generated during inflation, dark phase transitions and recombination in the early universe. 
Here we assume that PDMFs are produced from the so-called axion inflation~\cite{McAllister:2008hb}, in which a pseudo-scalar inflaton couples to the DP via an effective Chern-Simons term, $a\hat F_{\mu\nu} {\hat F}_{\sigma \tau} \varepsilon^{\mu\nu \sigma \tau } /(8\Lambda)$. Then, the PDMFs can be produced during inflation and can be kept until to today~\cite{Turner:1987bw,Garretson:1992vt,Anber:2006xt,Jimenez:2017cdr}.
The magneto hydrodynamic turbulence of these PDMFs may lead to the production of stochastic GWs~\cite{Caprini:2006jb}, which is interesting but beyond the scope of this study. 
Alternatively, the oscillations of ULDP in PDMFs in the early universe may also contribute to the SGWBs. In this section, we consider the relic density of stochastic GWs coming from this process. The power of the oscillation can be written as
\bea
{d \rho_{\rm GW}^{}  \over dt } = E_{\rm ULDP}^{} n_{\rm ULDP}^{} {\cal R}\left(\gamma^\prime \to {\rm SGW} \right)\;,
\eea 
where $E_{\rm ULDP}^{}$ and $n_{\rm ULDP}$ are the energy and the number density of the ULDP in early times, respectively. 
This formula is similar to that of GWs produced from cosmic string oscillations~\cite{Vachaspati:1984gt,Damour:2000wa}. 
Considering that the DP flux is homogeneous and isotropic, we need to average over the alignment of the DP velocity to the PDMF giving $\langle B_T^2\rangle =1/2 B^2$ with $B$ denoting the magnitude of the PDMF, in the oscillation formula.
The SGW energy density today is a redshifted total energy deposited from oscillations in early times
\bea
%\rho_{SGW}^{} (t_f, f) =\int_{t_i}^{t_f} {dt \over (1+z)^4} \cdot n \cdot P_{\rm SGW} (t, f^\prime ) {\partial f^\prime \over \partial f}
\rho_{\rm SGW}^{}=\int_{z_0}^{z_1} n_{\rm ULDP}(z_0) {\cal R}(z)  m_{\rm ULDP}^{} {dz \over (1+z)^2 H(z)} \;, 
\eea
where $n_{\rm ULDP}(z_0)$ is the number density of ULDP at the present time, $z$ is the redshift parameter, $z_1$ is the redshift when ULDP starts to oscillate, and $H(z)$ is the Hubble parameter. 
The relationship between the frequency of the  GW at production, $f^\prime$, and that at the present time, $f$, is $f^\prime = (1+z) f$. This formula divided by the critical energy density results in the relic abundance of SGW. Given that $\Omega_{\rm GW}(f) =  {d\rho_{\rm SGW} \over df }{f\over\rho_C}$, where $\rho_C=3 H_0^2 /(8 \pi G) $ being the critical density, one has
\bea
\Omega_{\rm GW} (f) = {n_{\rm ULDP}{\cal R}(f)\over H(f)} {f \over \rho_C}\;,
\eea
%\TL{$n_{\rm ULDP}$?}
where we have used the relationship between the frequency of the GW and the redshift: $f=m_{\rm ULDP}/ (1+z)$ to get the final formula.

\begin{figure}[t]
\centering
\includegraphics[width=0.45\textwidth]{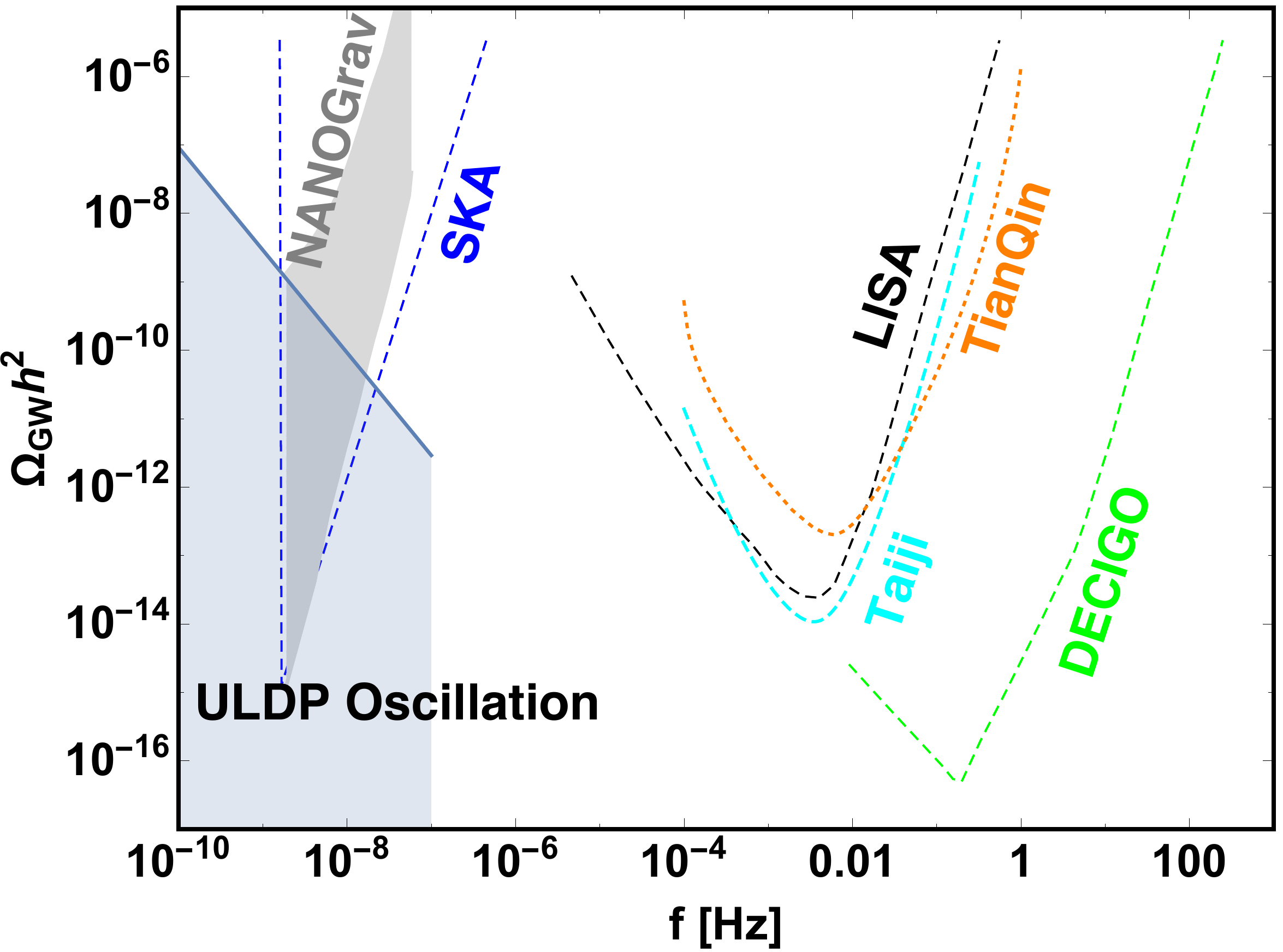}
\caption{The global relic abundance of the stochastic GWs produced via dark photon oscillations in the early universe as a function of the frequency (light blue region). The gray region represents the signal observed by the NANOGrav collaboration. The blue, orange, green, magenta and dark green dashed curves show the projected sensitivity of the SKA, LISA and DECIGO, TianQin and Taiji collaboration, respectively.} \label{fig:3GWRs}
\end{figure}

To estimate the relic density of the GW produced from the oscillation of ULDP in the early universe, one needs to first confirm the starting time of the oscillation. As mentioned above, PDMF can be produced via the axion inflation, however, there are various production mechanisms for the ULDP, such as  the so-called misalignment mechanism~\cite{Nelson:2011sf},  axion oscillation~\cite{Co:2018lka}, or the cosmic string decay~\cite{Long:2019lwl}, etc.  Considering that the ULDP produced from these mechanisms is always non-relativistic, the earlier the GW is generated, the lower its frequency is.  Since the GW with super-low frequency is out of the reach of current GW detectors, we only focus on the relic abundance of GW that may be detected today.  
In Fig.~\ref{fig:3GWRs}, we show the relic abundance of the stochastic GWs produced from the oscillations of the ULDP as the function of its frequency. We have taken $B(z_1)\leq1$ Gauss, which is red-shifted as $(1+z)^{-2}$~\cite{Fujita:2016igl}, $m_{\rm ULDP}=10^{-20}$ eV and  the cutoff of the redshift as $z_1= 10^{4}$, above which the GW from ULDP oscillation can not be detected. The light blue region represents the stochastic GWs produced via ULDP oscillation in the early universe. The gray region shows the signal observed by the NANOGrav collaboration, which apparently can be addressed by the ULDP oscillation mechanism. The observed GW band is caused by the redshift of the universe. If the ULDP is heavier, it may produce a GW that can be detected in the space-based or ground-based interferometer. It should be mentioned that the power law of the spectrum is mainly determined by the redshift behavior of the PDMF, a modification of which will change the shape of the spectrum.  

%\TL{*************** done here *****************}

%%%%%%%%%%%%%%%%%%%%%%%%%%%%%%%%%%
\section{Conclusion} 
\label{sec:Con}
%%%%%%%%%%%%%%%%%%%%%%%%%%%%%%%%%%

%GW are gradually becoming a new probe to the physics of  the early universe. Exploring the origin of stochastic GW beyond the conventional strategy is important. 

%The discovery of GWs opens a new window of exploring the physics of the early universe. Identifying sources of GWs in the early universe and their spectrums today turn to be  important tasks so as to guide the experimental detection of stochastic GWs. 
GW is becoming a new probe for the physics of the early universe. 
In this paper we have proposed a novel possibility of generating stochastic GWs via ULDP oscillations in the presence of PDMFs. 
For the first time, we have calculated the local energy density of these stochastic GWs in the galaxy, which arises  instantaneously from ULDP oscillations today. The local GW exhibits pulse-like spectrum with fixed frequency and can be detected in PTAs or future space-based interferometers.
We also find that the low-frequency GW signal observed by NANOGrav collaboration and other PTA experiments can be addressed by the oscillation of ULDP in the early universe. It should be mentioned that a pseudo-scalar ultralight DM, such as axion-like particles, may also oscillate into stochastic GWs in a similar way. This study has broadened the horizon of exploring the stochastic GWs.

\begin{acknowledgments}
W.~C. is supported by the National Natural Science Foundation of China (NSFC) (Grants No. 11775025 and No. 12175027). T.~L. is supported by the National Natural Science Foundation of China (Grant No. 12375096, 12035008, 11975129) and ``the Fundamental Research Funds for the Central Universities'', Nankai University (Grant No. 63196013).
\end{acknowledgments}

%%%%%%%%%%%%%%%%%%%%%%%%%%%%%%%%%%%%%
\bibliography{references}

\end{document}